\documentclass[aps,prl,twocolumn,superscriptaddress,showpacs,
nofootinbib,nobalancelastpage,nobibnotes]{revtex4}
\usepackage[dvips]{epsfig}
\usepackage{amsmath}

\begin{document}

\title{Decay of High-Energy Astrophysical Neutrinos}

\author{John F. Beacom}
\email{beacom@fnal.gov}

\author{Nicole F. Bell}
\email{nfb@fnal.gov}
\affiliation{NASA/Fermilab Astrophysics Center, Fermi National Accelerator
Laboratory, Batavia, Illinois 60510-0500}

\author{Dan Hooper}
\email{hooper@pheno.physics.wisc.edu}
\affiliation{Department of Physics, University of Wisconsin, 
Madison, Wisconsin 53706}

\author{Sandip Pakvasa}
\email{pakvasa@phys.hawaii.edu}
\affiliation{Department of Physics and Astronomy, University of Hawaii,
Honolulu, Hawaii 96822}
\affiliation{Theory Group, KEK, Tsukuba, Ibaraki 305-0801, Japan}

\author{Thomas J. Weiler}
\email{tom.weiler@vanderbilt.edu}
\affiliation{Department of Physics and Astronomy, Vanderbilt University,
Nashville, Tennessee 37235}

\date{November 19, 2002}

\begin{abstract}
Existing limits on the non-radiative decay of one neutrino to another
plus a massless particle (e.g., a singlet Majoron) are very weak.  The
best limits on the lifetime to mass ratio come from solar neutrino
observations, and are $\tau/m \agt 10^{-4}$ s/eV for the relevant mass
eigenstate(s).  For lifetimes even several orders of magnitude longer,
high-energy neutrinos from distant astrophysical sources would decay.
This would strongly alter the flavor ratios from the
$\phi_{\nu_e}:\phi_{\nu_{\mu}}:\phi_{\nu_{\tau}} = 1:1:1$ expected from
oscillations alone, and should be readily visible in the near future
in detectors such as IceCube.
\end{abstract}

\pacs{13.35.Hb, 14.60.Pq, 95.85.Ry \hspace{4cm} FERMILAB-Pub-02/243-A}

\maketitle


Neutrinos from astrophysical sources are expected to arise dominantly
from the decays of pions and their muon daughters, which results in
initial flavor ratios
$\phi_{\nu_e}:\phi_{\nu_{\mu}}:\phi_{\nu_{\tau}}$ of nearly $1:2:0$.
The fluxes of each mass eigenstate are given by $\phi_{\nu_i}=
\sum_{\alpha} \phi_{\nu_\alpha}^{\rm source} |U_{\alpha i}|^2$, where
$U_{\alpha i}$ are elements of the neutrino mixing matrix.  For three
active neutrino species (as we assume throughout) there is now strong
evidence to suggest that $\nu_{\mu}$ and $\nu_{\tau}$ are maximally
mixed and $U_{e3} \simeq 0$.  The consequent $\nu_\mu$--$\nu_\tau$
symmetry means that in the mass eigenstate basis the neutrinos are
produced in the ratios $\phi_{\nu_1}:\phi_{\nu_2}:\phi_{\nu_3} =
1:1:1$, independent of the solar mixing angle.  Oscillations do not
change these proportions, but only the relative phases between mass
eigenstates, which will be lost.  An incoherent mixture in the ratios
$1:1:1$ in the mass basis implies an equal mixture in any basis
(${\cal U} I {\cal U}^{\dagger} \equiv I$), and in particular the
flavor basis in which the neutrinos are detected~\cite{111}.  In this
{\it Letter} we show that neutrino decay could alter the measured
flavor ratios from the expected $1:1:1$ in a strong and distinctive
fashion.

We restrict our attention to two body decays
\begin{equation}
\nu_i \rightarrow \nu_j + X \;\;\; \rm{and} \;\;\;
\nu_i \rightarrow \overline{\nu}_j + X,
\end{equation}
where $\nu_i$ are neutrino mass eigenstates and $X$ denotes a very
light or massless particle, e.g. a Majoron.  Viable Majoron models 
which feature large neutrino decay rates have been discussed in 
Ref.~\cite{models}. We do not
consider either radiative two-body decay modes (which are constrained
by photon appearance searches to have very long lifetimes~\cite{RPP})
or three-body decays of the form $\nu \rightarrow \nu\nu\bar{\nu}$
(which are strongly constrained~\cite{sandip} by bounds on anomalous
$Z\nu\bar{\nu}$ couplings~\cite{santamaria}).  In contrast, the limits
on the decay modes considered here are very weak.  Beacom and Bell
have shown that the strongest reliable limit is $\tau/m \agt 10^{-4}$
s/eV, set by the solar neutrino data~\cite{BB}.  This limit is based
primarily on the non-distortion of the Super-Kamiokande spectrum~\cite{SK}, 
and takes into account the potentially competing distortions caused by
oscillations (see also Ref.~\cite{Choubey}) as well as the appearance
of active daughter neutrinos.  It is very likely that the SN 1987A
data place no limit at all on these neutrino decay modes, since decay
of the lightest mass eigenstate is kinematically forbidden, and even a
reasonable $\bar{\nu}_1$ flux alone can account for the data~\cite{frieman,BB}.

The strongest lifetime limit is thus too weak to eliminate the
possibility of astrophysical neutrino decay by a factor of about $10^7
\times (L/100 {\rm\ Mpc}) \times (10 {\rm\ TeV}/E)$~\cite{BB}.  A few
previous papers have considered aspects of the decay of high-energy
astrophysical neutrinos.  It has been noted that the disappearance of
all states except $\nu_1$ would prepare a beam that could in principle
be used to measure elements of the neutrino mixing matrix, namely the
ratios $U^2_{e1}:U^2_{\mu 1}:U^2_{\tau 1}$~\cite{farzan}.  The
possibility of measuring neutrino lifetimes over long baselines was
mentioned in Ref.~\cite{Weiler}, and some predictions for decay in
four-neutrino models were given in Ref.~\cite{Keranen}.  
We will show that the particular values and small uncertainties on
the neutrino mixing parameters allow for the first time very
distinctive signatures of the effects of neutrino decay on the
detected flavor ratios.  The expected increase in neutrino lifetime
sensitivity (and corresponding anomalous neutrino couplings) by
several orders of magnitude makes for a very interesting test of
physics beyond the Standard Model; a discovery would mean physics much
more exotic than neutrino mass and mixing alone.  We will show that
neutrino decay cannot be mimicked by either different
neutrino flavor ratios at the source or other non-standard neutrino
interactions.

A characteristic feature of decay is its strong energy dependence:
$\exp\left(-L/\tau_{\rm lab}\right) = \exp\left(-Lm/E\tau\right)$,
where $\tau$ is the rest-frame lifetime.  However, we will assume
that decays are always complete, i.e., that these exponential factors
vanish.  This is reasonable because there is a minimum $L/E$
value set by the shortest distances (typically hundreds of Mpc) and
the maximum energies that will be visible in a given detector (the
spectra considered are steeply falling).  The assumption of complete
decay means we do not have to consider the distance and intensity
distributions of sources.  We assume an isotropic diffuse flux
of high-energy astrophysical neutrinos, and can thus neglect the
angular deflection of daughter neutrinos from the trajectories of
their parents~\cite{lindner}.  It is uncertain if astrophysical
sources produce the same numbers of neutrinos and antineutrinos.  
Though the detectors cannot distinguish
neutrinos from antineutrinos, their cross sections are different,
and this could cause confusion in the deduced flavor ratios.  However,
the antineutrino-neutrino cross section ratio is 0.7 at 10 TeV, and 
rapidly approaches unity at higher energies.

{\bf Disappearance only.---}
We first assume that there are no detectable decay products, that is, 
the neutrinos simply disappear.  This limit is interesting for decay to
`invisible' daughters, such as a sterile neutrino, and also for decay to 
active daughters if the source spectrum falls sufficiently steeply with 
energy.  In the latter case, the flux of daughters of degraded energy 
may make a negligible contribution to the total flux at a given energy.
Since coherence will be lost we have
\begin{eqnarray}
\label{simple1}
\phi_{\nu_\alpha}(E) &=& \sum_{i\beta} \phi^{\rm source}_{\nu_\beta}(E)
|U_{\beta i}|^2 |U_{\alpha i}|^2 e^{-L/\tau_i(E)} \\
\label{simple}
&\xrightarrow{L\gg \tau_i}&  
\sum_{i(stable), \beta} \phi^{\rm source}_{\nu_\beta}(E)
|U_{\beta i}|^2 |U_{\alpha i}|^2,
\end{eqnarray}
where the $\phi_{\nu_\alpha}$ are the fluxes of $\nu_{\alpha}$,
$U_{\alpha i}$ are elements of the neutrino mixing matrix and $\tau_i$
are the neutrino lifetimes in the laboratory frame.
Eq.~(\ref{simple}) corresponds to the case where decay is complete by
the time the neutrinos reach Earth, so only the stable states are
included in the sum.

The simplest case (and the most generic expectation) is a normal
hierarchy in which both $\nu_3$ and $\nu_2$ decay, leaving only the
lightest stable eigenstate $\nu_1$.  In this case the flavor ratio is
$U_{e1}^2:U_{\mu1}^2:U_{\tau1}^2$~\cite{farzan}.  Thus if $U_{e3}=0$
\begin{equation} 
\label{generic}
\phi_{\nu_e}:\phi_{\nu_\mu}:\phi_{\nu_\tau} =
\cos^2 \theta_{\odot}:\frac{1}{2}\sin^2 \theta_{\odot}:
\frac{1}{2}\sin^2 \theta_{\odot} \simeq 6:1:1,
\end{equation}
where $\theta_{\odot}$ is the solar neutrino mixing angle, which we
have set to $30^{\circ}$.  Note that this is an extreme deviation of
the flavor ratio from that in the absence of decays.  It is difficult
to imagine other mechanisms that would lead to such a high ratio of
$\nu_e$ to $\nu_{\mu}$.  
Here and throughout we concentrate on the flavor ratios, since the original
source fluxes are unknown.
In the case of an inverted hierarchy, $\nu_3$ is the lightest and
hence stable state, and so
\begin{equation}
\label{inverted}
\phi_{\nu_e}:\phi_{\nu_\mu}:\phi_{\nu_\tau} =
U_{e3}^2:U_{\mu3}^2:U_{\tau3}^2 = 0:1:1.
\end{equation}

\begin{figure}[t]
\begin{center}
\epsfig{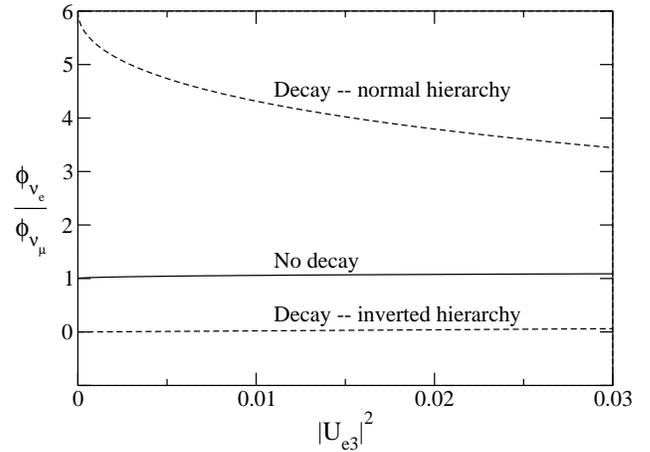}
\caption{\label{ratio} The effect of the presently unknown $U_{e3}$ on
the $\phi_{\nu_e}/\phi_{\nu_\mu}$ ratio.  We have fixed
$\theta_{\odot} = 30^{\circ}$ and $\theta_{\rm atm} = 45^{\circ}$.
Although varying these angles affects the flux ratios to a similar
extent as $U_{e3}$, they will be precisely
measured in the near future.  In all cases, the three scenarios are
very distinct.}
\end{center}
\end{figure}

If $U_{e3}=0$ and $\theta_{\rm atm}=45^\circ$, each mass eigenstate
has equal $\nu_{\mu}$ and $\nu_{\tau}$ components.  Therefore, decay
cannot break the equality between the $\phi_{\nu_\mu}$ and
$\phi_{\nu_\tau}$ fluxes and thus the $\phi_{\nu_e}:\phi_{\nu_\mu}$ 
ratio contains
all the useful information.  The variation of the
$\phi_{\nu_e}:\phi_{\nu_\mu}$ ratio with non-zero $U_{e3}$ (up to the
maximum allowed value, $|U_{e 3}|^2 \alt 0.03$~\cite{Ue3}) is shown in
Fig.~\ref{ratio}.  In the no-decay case, the variation from $1:1:1$ is
negligibly small.  While the relative effect can be larger if neutrino
decay occurs, the three cases shown are always quite distinct.  In
addition, the ratio of the $\nu_\mu$ and $\nu_\tau$ components can
also change, e.g., Eq.~(\ref{generic}) could be as
extreme as $U_{e1}^2:U_{\mu1}^2:U_{\tau1}^2 = 3.5 : 1: 0.3$.
Hereafter, we set $U_{e3} = 0$.

{\bf Appearance of daughter neutrinos.---}
If neutrino masses are quasi-degenerate, the daughter neutrino carries
nearly the full energy of the parent.  An interesting and convenient
feature of this case is that we can treat the effects of the daughters
without making any assumptions about the source spectra.
Including daughters of full energy, we have 
\begin{eqnarray}
\phi_{\alpha}(E) 
&\xrightarrow{L\gg \tau_i}&
\sum_{i\beta} \phi^{\rm source}_{\beta}(E)
|U_{\beta i}|^2 |U_{\alpha i}|^2  \nonumber \\
&+& \sum_{ij\beta} \phi^{\rm source}_{\beta}(E) 
|U_{\beta j}|^2 |U_{\alpha i}|^2 B_{j \rightarrow i}  
\label{degen}
\end{eqnarray}
where $B$ is a branching fraction and stable and unstable states are
denoted henceforth by $i$ and $j$ respectively.

If instead the neutrino mass spectrum is hierarchical, the daughter
neutrinos will be degraded in energy with respect to the parent, so
that
\begin{eqnarray}
&\phi_{\nu_\alpha}(E) &
\xrightarrow{L\gg \tau_i}
\sum_{i\beta} \phi^{\rm source}_{\nu_\beta}(E)
|U_{\beta i}|^2 |U_{\alpha i}|^2   \\
&+& \int_E^\infty \! dE' \; W_{E'E} \sum_{ij\beta} 
\phi^{\rm source}_{\nu_\beta}(E') 
|U_{\beta j}|^2 |U_{\alpha i}|^2 B_{j \rightarrow i}, \nonumber
\end{eqnarray}
where $E$ is the daughter and $E'$ is the parent energy.  The
normalized energy spectrum of the daughter is given by
\begin{equation}
W_{E'E}=\frac{1}{\Gamma(E')}\frac{d\Gamma(E',E)}{dE}.
\end{equation} 

If the neutrinos are Majorana particles, daughters of both helicities
will be detectable (as neutrinos or antineutrinos), whereas if they
are Dirac particles, daughters of one helicity will be sterile and
hence undetectable.  In the rest frame of the parent neutrino, the
angular distributions for decays which conserve and flip helicity are
proportional to $\cos^2(\theta^*/2)$ and $\sin^2(\theta^*/2)$
respectively, where $\theta^*$ is the angle of the daughter neutrino
with respect to the (lab frame) momentum of the parent.
In the limit $m_{\rm daughter} \ll m_{\rm parent}$, the corresponding
energy distributions in the lab frame are $E/E'^2$ and $(E'-E)/E'^2$.

In the case of Majorana neutrinos, we may drop the distinction between 
neutrino and antineutrino daughters and sum over helicities.
Assuming the source spectrum to be a simple power
law, $E^{-\alpha}$, we find
\begin{eqnarray}
\label{degrade}
\phi_{\nu_\alpha}(E) 
&\xrightarrow{L\gg \tau_i}&
\sum_{i \beta} \phi^{\rm source}_{\nu_\beta}(E)
|U_{\beta i}|^2 |U_{\alpha i}|^2 \nonumber \\
&+& \frac{1}{\alpha} \sum_{ij\beta} \phi^{\rm source}_{\nu_\beta}(E) 
|U_{\beta j}|^2 |U_{\alpha i}|^2 B_{j \rightarrow i}
\end{eqnarray}
This is identical to the expression in Eq.~(\ref{degen}) except 
for the overall factor of $1/\alpha$ in front of the second term.
For Dirac neutrinos we detect only the daughters that conserve 
helicity, the effect of which is only to change the numerical 
coefficient of the second sum in Eq.~(\ref{degrade}). 
Thus, although the flavor ratio will differ from the cases above, 
it is still independent of energy---i.e., decay does not introduce 
a spectral distortion of the power law.
We stress that we have assumed a simple but reasonable power law
spectrum $E^{-\alpha}$; a broken power law spectrum, e.g., would lead
to a more complicated energy dependence.

{\bf Uniqueness of decay signatures.---}
Depending on which of the mass eigenstates are unstable, the decay
branching ratios, and the hierarchy of the neutrino mass eigenstates,
quite different ratios result.
For the normal hierarchy, some possibilities are shown in Table~I.

\begin{table}[ht]
\label{ratios}
\caption{Flavor ratios for various decay scenarios.}
\begin{tabular}{c|l|l|c}
\hline\hline
Unstable & Daughters & Branchings &  
$\phi_{\nu_e}:\phi_{\nu_\mu}:\phi_{\nu_\tau}$ \\
\hline\hline
$\nu_2$, $\nu_3$  & anything & irrelevant & $6:1:1$ \\
\hline
$\nu_3$	          & sterile  & irrelevant & $2:1:1$ \\
\hline
$\nu_3$          & full energy & $B_{3 \rightarrow 2}=1$  & $1.4:1:1$ \\
                 & degraded ($\alpha=2$)               &  & $1.6:1:1$ \\
\hline
$\nu_3$          & full energy & $B_{3 \rightarrow 1}=1$   & $2.8:1:1$ \\  
                 & degraded ($\alpha=2$)               &   & $2.4:1:1$ \\
\hline
$\nu_3$          & anything & $B_{3 \rightarrow 1}=0.5$    & $2:1:1$ \\
                 &  &  $B_{3 \rightarrow 2}=0.5$           & \\  
\hline\hline  
\end{tabular}
\end{table}

The most natural possibility with unstable neutrinos is that the
heaviest two mass eigenstates both completely decay.  The resulting
flavor ratio is just that of the lightest mass
eigenstate, independent of energy and whether daughters are detected 
or not.  For normal and inverted hierarchies we obtained  
$6:1:1$ and $0:1:1$ respectively.  Interestingly, both cases have 
extreme $\phi_{\nu_e}:\phi_{\nu_\mu}$ ratios, which provides a
very useful diagnostic.  Assuming no new physics besides decay, a
ratio greater than 1 suggests the normal hierarchy, while a ratio
smaller than 1 suggests an inverted hierarchy.  In the case that
decays are not complete these trends still hold, even though the
limits of Eqs.~(\ref{generic},\ref{inverted}) would not be reached.
The case of incomplete decay might be identified by measuring
different flux ratios in different energy ranges.  It is interesting
to note that complete decay cannot reproduce $1:1:1$.  One of the mass
eigenstates does have a flavor ratio similar to $1:1:1$, but it is the
heavier of the two solar states and cannot be the lightest,
stable state.  (A possible but unnatural exception occurs if only this
state decays).

An important issue is how unique decay signatures would be.  
Are there other scenarios (either non-standard astrophysics or
neutrino properties) that would give similar ratios? 
There exist astrophysical neutrino production models with different 
initial flavor ratios, such as $0:1:0$~\cite{Rachen}, for which the 
detected flavor ratios (in the absence of decay) would be about $0.5:1:1$.
However, since the mixing angles $\theta_\odot$ and $\theta_{\rm atm}$ 
are both large, and since the neutrinos are produced and detected in 
flavor states, no initial flavor ratio can result in a measured 
$\phi_{\nu_e}:\phi_{\nu_\mu}$ ratio anything like that of our two main cases, 
$6:1:1$ and $0:1:1$.

In terms of non-standard particle physics, decay is unique in the sense 
that it is ``one-way'', unlike, say, oscillations or magnetic moment
transitions.  Since the initial flux ratio in the mass basis is $1:1:1$, 
magnetic moment transitions between (Majorana) mass eigenstates cannot 
alter this ratio, due to the symmetry between $i\rightarrow j$ and 
$j \rightarrow i$ transitions.
On the other hand, if neutrinos have Dirac masses, magnetic moment transitions
(both diagonal and off-diagonal) turn active neutrinos into sterile states, so
the same symmetry is not present.  However, the process will not be complete
in the same way as decay---it will average out at 1/2, so there is no way 
we could be left with a only single mass eigenstate.


{\bf Experimental Detectability.---}
Deviations of the flavor ratios from $1:1:1$ due to possible decays 
are so extreme that they should be readily identifiable~\cite{reviews}.
Upcoming high energy neutrino experiments, such as IceCube~\cite{icecube}, will
not have perfect abilities to separately measure the neutrino flux in each
flavor.  However, the quantities we need are closely related to the
observables, in particular in the limit of $\nu_\mu$--$\nu_\tau$ symmetry
($\theta_{\rm atm} = 45^\circ$ and $U_{e3} = 0$), in which all mass
eigenstates contain equal fractions of $\nu_\mu$ and $\nu_\tau$.  In that
limit, the fluxes for $\nu_\mu$ and $\nu_\tau$ are always in the ratio
$1:1$, with or without decay.  This is useful since the $\nu_\tau$ flux is
the hardest to measure.

Detectors such as IceCube will be able to directly measure the $\nu_\mu$
flux by long-ranging muons which leave tracks through the
detector.  The charged-current interactions of $\nu_e$ produce
electromagnetic showers.  However, these may be hard to
distinguish from hadronic showers caused by all flavors through their
neutral-current interactions, or from the charged-current interactions of
$\nu_\tau$ (an initial hadronic shower followed by either an
electromagnetic or hadronic shower from the tau lepton decay)~\cite{showers}.
We thus consider our only
experimental information to be the number of muon tracks and the number of
showers.

The relative number of shower events to track events can be related to the
most interesting quantity for testing decay scenarios, i.e., the $\nu_e$
to $\nu_\mu$ ratio.  The precision of the upcoming experiments should be
good enough to test such extreme flavor ratios produced by decays. If
electromagnetic and hadronic showers can be separated, then the precision
will be even better.

Comparing, for example, the standard flavor ratios of $1:1:1$ to
the possible $6:1:1$ generated by decay, 
the more numerous electron neutrino flux
will result in a substantial increase in the number of showers compared to
the number of muon events.  The details of this observation depends on the
range of muons generated in or around the detector and the ratio of
charged to neutrino current cross sections.  This measurement will be
limited by the energy resolution of the detector and the ability to 
reduce the atmospheric neutrino background.  The atmospheric
background drops rapidly with energy and should be negligibly small above
the PeV scale.


{\bf Discussion and Conclusions.---}
We have presented our results above in terms of the ratios of fluxes in
each neutrino flavor.  These ratios are energy-independent because we have
assumed that the ratios at production are energy-independent, that all
oscillations are averaged out, and that all possible decays are complete.  
The first two assumptions are rather generic, and the third is a
reasonable simplifying assumption.  In the standard scenario with only 
oscillations, the final flux ratios are
$\phi_{\nu_e}:\phi_{\nu_{\mu}}:\phi_{\nu_{\tau}} = 1:1:1$.  In the
cases with decay, we have shown rather different possible flux ratios, for
example $6:1:1$ in the normal hierarchy and $0:1:1$ in the inverted
hierarchy.  These deviations from $1:1:1$ are so extreme that they should
be readily measurable.

These clear and striking predictions for the effects of neutrino decay
on the measured flavor ratios depend strongly on recent progress in
measuring neutrino mixing parameters.  In particular, it is very
significant that $\theta_\odot \simeq 30^\circ$~\cite{SNO} is well
below the maximal $45^\circ$, for which Eq.~(\ref{generic}) would
instead be a much less dramatic $2:1:1$.  In addition, $\theta_\odot <
45^\circ$ means that $\delta m^2_{12} > 0$ and hence that $\nu_2$
(with flavor ratios $0.7:1:1$) can never be the lightest mass
eigenstate.  Maximal $\theta_{\rm atm}$ and very small
$U_{e3}$ also make the predictions clearer.  The hierarchy of $\nu_3$
relative to the two solar states is unknown, but in either case neutrino
decay will be stringently tested by upcoming measurements
of astrophysical neutrinos.

\bigskip
We thank B. Kayser and J. Learned for
illuminating discussions.  J.F.B. and N.F.B. were supported by
Fermilab (under DOE contract DE-AC02-76CH03000) and by
NASA grant NAG5-10842, D.H by the 
Wisconsin Alumni Research Foundation and by DOE grant DE-FG02-95ER40896,
S.P. by DOE grant DE-FG03-94ER40833, and T.W. by DOE grant DE-FG05-85ER40226.



\end{document}